\documentclass[journal=tches,final]{iacrtrans}

\setfirstpage{0}
\setlastpage{0}
\setvolume{0}
\setnumber{0}

\setISSN{2569-2925}
\makeatletter
\setDOI{10.46586/tches.v\IACR@vol.i\IACR@no.\IACR@fp-\IACR@lp}
\makeatother

\setkeys{IACR}{journal=tches}

\setvolume{2026}
\setnumber{3}
\setfirstpage{465}
\setlastpage{488}

\usepackage{multirow}
\usepackage{booktabs} 
\usepackage{tabularx}
\usepackage{subcaption} 
\usepackage{graphicx} 

\usepackage{verbatim} 

\usepackage{tikz}
\usetikzlibrary{calc,positioning,arrows.meta}
\usetikzlibrary{arrows}

\newcommand{\bak}{\texttt{BAKSHEESH}}
\newcommand{\de}{\texttt{DEFAULT}}
\newcommand{\dl}{\texttt{DEFAULT-LAYER}}
\newcommand{\dc}{\texttt{DEFAULT-CORE}}
\newcommand{\z}{\(\cdot\)}

\author{Hanbeom Shin\inst{1} \and Insung Kim\inst{1} \and Sunyeop Kim\inst{1,2} \and Byoungjin Seok\inst{3} \and Deukjo Hong\inst{4} \and Jaechul Sung\inst{5} \and Seokhie Hong\inst{6} \and Sangjin Lee\inst{1} \and Dongjae Lee\footnote{Corresponding Author}\inst{7}}
\authorrunning{Shin et al.}
\institute{
Korea University, South Korea, \email{{newonetiger,cmcom35,sangjin}@korea.ac.kr}
\and 
Nanyang Technological University, Singapore, \email{kin3548@gmail.com}
\and 
Hansung University, South Korea, \email{bjseok@hansung.ac.kr}
\and 
Jeonbuk National University, South Korea, \email{deukjo.hong@jbnu.ac.kr}
\and 
University of Seoul, South Korea, \email{jcsung@uos.ac.kr}

\and
SmartM2M, South Korea, \email{shhong@smartm2m.co.kr}
\and
Kangwon National University, South Korea, \email{dongjae.lee@kangwon.ac.kr} 
}

\title{MIFA: An MILP-based Framework for Improving Differential Fault Attacks}

\begin{document}

\maketitle


\begin{abstract}
At ASIACRYPT 2021, Baksi et al. introduced \de{}, a block cipher designed to algorithmically resist Differential Fault Attack (DFA), claiming 64-bit DFA security regardless of the number of injected faults. At EUROCRYPT 2022, Nageler et al. demonstrated that \de{}'s claimed DFA resistance can be broken by applying an information-combining technique. More recently, at ASIACRYPT 2024, Jana et al. improved DFA by searching for differential trails with a single solution. They showed that, for \de{} with a simple key schedule, injecting five faults at the fifth-to-last round reduces the key space to one, and for \bak{}, injecting twelve faults at the third-to-last round achieves the same result. 
In this paper, we propose a new DFA framework that utilizes a Mixed-Integer Linear Programming (MILP) solver. This framework makes it possible to attack deeper rounds than previously achieved, reducing the number of fault injections required for key recovery. Furthermore, we present a method to determine the most efficient fault injection bit positions by systematically analyzing the input differences from all possible single bit-flip faults, thereby further reducing the required number of faults. This systematic analysis has the significant advantage of allowing us to theoretically calculate the required number of faults. Applying our framework, for \de{}, injecting three faults at the sixth-to-last round and two faults at the seventh- and eighth-to-last rounds reduces the key space to one.
\end{abstract}

\section{Introduction}
\label{sec:introduction}

A fault attack is a class of physical attacks in which intentional faults are injected during the execution of a cryptographic or computing device, and the resulting faulty outputs or abnormal behaviors are exploited to extract secret information, recover internal states, or break security mechanisms. Faults can be induced through methods such as voltage or clock glitches \cite{o2016fault}, electromagnetic pulses \cite{moro2013electromagnetic}, laser injections \cite{selmke2015precise}, or memory disturbances \cite{kim2014flipping}, making fault attacks a powerful and practical threat to both cryptographic implementations and general-purpose systems.

Induced faults are commonly categorized into several models depending on how they alter data and computations. A bit-flip fault inverts one or more bits of a register or memory value. A zeroing fault (or stuck-at-zero) forces the affected value to zero, while a stuck-at-one fault fixes bits to one regardless of their intended value. A bit-set fault sets a specific bit to a desired value, either 0 or 1. A randomizing fault replaces the value with an unpredictable state. In addition to these, an instruction skip fault occurs when an operation in the execution flow is bypassed. These models describe typical manifestations of physical fault injection and are widely used in the evaluation of cryptographic implementations.

In this work, we primarily focus on the bit-flip fault model. Achieving such precision is feasible in practice~\cite{barenghi2010fault}. According to previous studies, techniques such as laser fault injection offer high accuracy in both space and time, enabling the induction of precise single bit-flip faults in the cryptographic state~\cite{agoyan2010how, dutertre2012fault, selmke2015precise}. Furthermore, electromagnetic (EM) fault injection serves as an alternative method that does not require de-packaging the chip, and practical implementations of precise bit-level fault injections via EM have been demonstrated~\cite{ordas2015injection}. These advancements suggest that the bit-flip fault model is not merely a theoretical assumption but a practical threat.

Once a specific fault model is justified, the subsequent propagation of the fault within the cipher is determined solely by the cipher's design specifications and internal operations. In other words, the faulty ciphertexts obtained from a physical device are mathematically identical to those produced by a simulation that accurately implements the fault model. Consequently, software simulations serve as a sufficient and robust means to verify the attack's efficacy, including the required number of faults and computational complexity. This validation approach is consistent with methodologies employed in recent cryptanalytic studies~\cite{nageler2022information, jana2024more}.

Research on fault attacks in cryptography began with the introduction of Differential Fault Attack (DFA) by Biham and Shamir at CRYPTO 1997, which demonstrated that secret keys could be recovered by comparing correct and faulty outputs~\cite{biham1997differential}. Biham et al. presented Impossible Differential Fault Analysis (IDFA) at FSE 2005, which eliminates incorrect key candidates that produce an impossible differential \cite{biham2005impossible}. Clavier proposed Ineffective Fault Analysis (IFA) at CHES 2007, showing that even faults that cause no visible output difference can still leak key information~\cite{clavier2007secret}. Later, Fuhr et al. introduced Statistical Fault Analysis (SFA) at FDTC 2013, which recovers keys by collecting only faulty ciphertexts and exploiting statistical biases without requiring any correct outputs~\cite{fuhr2013fault}. Zhang et al. presented Persistent Fault Analysis (PFA) at CHES 2018, where persistent faults are injected into constants such as S-boxes and later exploited through statistical analysis of collected faulty ciphertexts~\cite{zhang2018persistent}. In the same year, Dobraunig et al. proposed Statistical Ineffective Fault Attack (SIFA) at CHES 2018, which leverages the probability distribution of ineffective outputs for key recovery~\cite{dobraunig2018sifa}. More recently, at ACNS 2023, Kundu et al. presented Division property based Fault Analysis (DiFA)~\cite{kundu2023divide}. Subsequently, Jana et al. introduced Statistical Differential Fault Analysis (SDFA) at ASIACRYPT 2024, combining differential analysis with statistical techniques~\cite{jana2024more}, and Kundu et al. proposed ToFA (Truncated impossible differential based Fault Analysis) at CHES 2025, which applies truncated impossible differentials to fault analysis~\cite{kundu2025tofa}. Various fault attacks on cryptographic systems have been proposed over the years and continue to be an active area of research.

DFA injects faults at specific points during the execution of a cryptographic algorithm and recovers secret keys by analyzing the differences between correct and faulty outputs. The attacker typically injects single or a few faults into intermediate states after several rounds and analyzes how the injected faults propagate through nonlinear components such as S-boxes to infer internal states and key bits. DFA is highly powerful because it can extract a large amount of information with only a small number of faults, but it requires pairs of correct and faulty outputs and precise control over the timing and location of fault injection to succeed. In block cipher settings, performing DFA on rounds that are farther from the plaintext and ciphertext, namely closer to the middle rounds, is considered more challenging from a cryptanalytic perspective, and attacks that succeed with fewer injected faults are regarded as more valuable.

From a defense perspective, redundancy-based countermeasures are effective but often incur significant implementation overheads in terms of area and power consumption~\cite{kundu2025tofa}. Specifically, this approach involves duplicating intermediate states to perform redundant computations for the remaining rounds and suppressing the output if a mismatch between the resulting ciphertexts is detected, thereby preventing the leakage of faulty outputs induced within the protected region. To balance security and efficiency, designers typically limit these protections to the final rounds, which are traditionally perceived as the most vulnerable to DFA~\cite{kundu2025tofa}. In this context, successful attacks on deeper rounds carry significant weight as they compel designers to extend the scope of these expensive countermeasures. For example, consider an 80-round cipher protected against a known 5-round DFA by applying countermeasures to the last 5 rounds. If a new 8-round DFA is introduced, it effectively bypasses the existing protection, rendering the cipher vulnerable. To defend against this new threat, the protected region must be expanded, which increases the implementation overhead ratio from roughly $5/80$ to $8/80$. Consequently, fault attacks that succeed on deeper rounds, specifically those closer to the middle of the cipher, pose a significant practical threat by forcing an increase in defense costs.

\de{}, proposed by Baksi et al. at ASIACRYPT 2021, aims to provide algorithmic DFA resistance using S-boxes with Linear Structures (LS), claiming that the number of key candidates remains at least $2^{64}$ regardless of the number of injected faults~\cite{baksi2022default}. However, this claim was refuted by subsequent studies. Nageler et al. first demonstrated at EUROCRYPT 2022 that the security could be significantly undermined using an information-combining technique~\cite{nageler2022information}. Subsequently, Jana et al. presented a more advanced attack at ASIACRYPT 2024 using differential trails (i.e., specific propagation trails of differences through the cipher's internal rounds) with a single solution under the bit-flip fault model~\cite{jana2024more}. They successfully executed a 5-round DFA, recovering the unique key with only 5 faults for the simple key schedule and $36 + \alpha$ faults for the rotating key schedule. These findings revealed that the originally claimed $2^{64}$-level DFA security of \de{} does not actually hold.

\subsection*{Our Contributions}

Building upon the work of Jana et al.~\cite{jana2024more}, we propose a new DFA framework, MIFA (MILP-based Differential Fault Attack), which significantly improves their approach under the bit-flip fault model. MIFA systematically searches for differential trails with a single solution targeting deeper rounds and maximizes attack efficiency. The framework is illustrated in Figure~\ref{fig:framework_diagram}, and our main contributions are as follows:

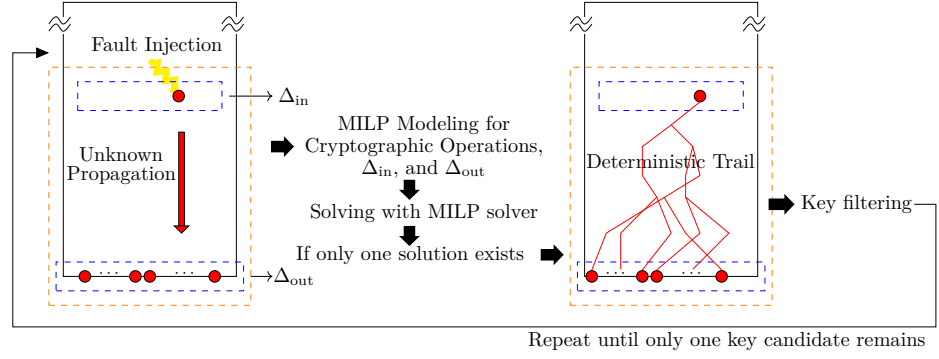
\begin{figure}[htbp]
    \centering
    \resizebox{\textwidth}{!}{%
        \usetikzlibrary{decorations.pathmorphing}
\usetikzlibrary{arrows.meta}
\begin{tikzpicture}

\begin{scope} [xshift=0, yshift=0]
  \draw[line width=3pt, yellow, decorate, decoration={zigzag, segment length=6pt, amplitude=2pt}]
    (-4,0.125) -- (-4.5,0.625);
\draw  (-6,1.625) rectangle (-3,-3.125);
\fill [white]  (-6.25,1.375) rectangle (-2.75,1.0875);

\draw  plot[smooth, tension=.7] coordinates {(-5.875,1.375) (-6,1.25) (-6.125,1.375) (-6.25,1.25)};
\draw  plot[smooth, tension=.7] coordinates {(-5.875,1.25) (-6,1.125) (-6.125,1.25) (-6.25,1.125)};

\draw  plot[smooth, tension=.7] coordinates {(-2.875,1.375) (-3,1.25) (-3.125,1.375) (-3.25,1.25)};
\draw  plot[smooth, tension=.7] coordinates {(-2.875,1.25) (-3,1.125) (-3.125,1.25) (-3.25,1.125)};

\draw  [fill=red](-4,0) circle (0.1cm);
\draw  [fill=red](-3.375,-3.125) circle (0.1cm);
\draw  [fill=red](-4.5,-3.125) circle (0.1cm);
\draw  [fill=red](-4.75,-3.125) circle (0.1cm);
\draw  [fill=red](-5.625,-3.125) circle (0.1cm);
\draw [fill=red] (-4,-0.625) rectangle (-3.9,-2.25);
\draw [fill =red] (-4.075,-2.25) -- (-3.95,-2.375) -- (-3.825,-2.25) -- cycle;
\draw [dashed, blue]  (-5.75,0.25) rectangle (-3.25,-0.25);
\draw [dashed, orange]  (-6.25,0.5) rectangle (-2.75,-3.625);
\draw [dashed, blue]  (-6.125,-2.875) rectangle (-2.875,-3.375);
\node at (-5.1875,-3.0625) {\footnotesize{$\cdots$}};
\node at (-3.875,-3.0625) {\footnotesize{$\cdots$}};
\end{scope}

\begin{scope} [xshift=250, yshift=0]
\draw  (-5.75,1.625) rectangle (-2.75,-3.125);
\fill [white]  (-6,1.375) rectangle (-2.5,1.0875);
\draw  plot[smooth, tension=.7] coordinates {(-5.625,1.375) (-5.75,1.25) (-5.875,1.375) (-6,1.25)};
\draw  plot[smooth, tension=.7] coordinates {(-5.625,1.25) (-5.75,1.125) (-5.875,1.25) (-6,1.125)};

\draw  plot[smooth, tension=.7] coordinates {(-2.625,1.375) (-2.75,1.25) (-2.875,1.375) (-3,1.25)};
\draw  plot[smooth, tension=.7] coordinates {(-2.625,1.25) (-2.75,1.125) (-2.875,1.25) (-3,1.125)};

\draw  [fill=red](-3.75,0) circle (0.1cm);
\draw  [fill=red](-3.375,-3.125) circle (0.1cm);
\draw  [fill=red](-4.5,-3.125) circle (0.1cm);
\draw  [fill=red](-4.75,-3.125) circle (0.1cm);
\draw  [fill=red](-5.625,-3.125) circle (0.1cm);
\draw [dashed, blue]  (-5.5,0.25) rectangle (-3,-0.25);
\draw [dashed, orange]  (-6,0.5) rectangle (-2.5,-3.625);
\draw [dashed, blue]  (-5.875,-2.875) rectangle (-2.625,-3.375);

\draw [red] (-3.75,-0.1) -- (-4.25,-0.5) -- (-4.75,-0.875) -- (-4.75,-1.375) -- (-4.5,-1.75) -- (-5.125,-2.375) -- (-5.125,-3);
\draw [red] (-4.25,-0.5) -- (-3.875,-0.875) -- (-4,-1.375) -- (-4,-1.75) -- (-3.75,-2.375) -- (-4.5,-3.0125);
\draw [red] (-3.875,-0.875) -- (-3.75,-1.375) -- (-5.375,-2.375) -- (-5.625,-3.0125);
\draw [red] (-4,-1.75) -- (-3.25,-2.375) -- (-3.875,-3);
\draw [red] (-4.5,-1.75) -- (-4.25,-2.375) -- (-4.75,-3.0125);
\draw [red] (-4.375,-1.75) -- (-4,-2.375) -- (-3.375,-3.0125);
\node at (-5.1875,-3.0625) {\footnotesize{$\cdots$}};
\node at (-3.875,-3.0625) {\footnotesize{$\cdots$}};
\end{scope}

\draw [->] (-3.125,0) -- (-2.375,0);
\node at (-2,0) {$\Delta_{\textnormal{in}}$};
\draw [->] (-2.75,-3.125) -- (-2.375,-3.125);
\node at (-2,-3.125) {$\Delta_{\textnormal{out}}$};
\node at (-5,-1) {{Unknown}};
\node at (-5,-1.375) {{Propagation}};
\node at (-4.375,0.875) {{Fault Injection}};
\node at (4.525,-1.125) {{Deterministic Trail}};

\node at (0.25,-0.5) {{MILP Modeling for}};
\node at (0.25,-0.875) {{Cryptographic Operations,}};
\node at (0.25,-1.25) {{$\Delta_{\text{in}}$, and $\Delta_{\text{out}}$}};

\node at (0.25,-2.025) {{Solving with MILP solver}};
\node at (0.0125,-2.75) {{If only one solution exists}};
\draw[-{Triangle[length=5pt,width=10pt]}, line width=2mm] (-2.4,-0.875) -- (-2,-0.875);
\draw[-{Triangle[length=5pt,width=10pt]}, line width=2mm] (0,-1.45) -- (0,-1.8);
\draw[-{Triangle[length=5pt,width=10pt]}, line width=2mm] (0,-2.25) -- (0,-2.6);
\draw[-{Triangle[length=5pt,width=10pt]}, line width=2mm] (0.275+2,0.175-2.925) -- (0.275+2.4,0.175-2.925);
\draw[-{Triangle[length=5pt,width=10pt]}, line width=2mm] (6.35-0.125,-1.875) -- (6.75-0.075,-1.875);
\node at (7.75,-1.875) {{Key filtering}};
\node at (5.5,-4.25) {{Repeat until only one key candidate remains}};
\draw [-triangle 45] (8.75,-1.875) -- (9.125,-1.875) -- (9.125,-4) -- (-6.875,-4) -- (-6.875,0.75) -- (-6.25,0.75);

\end{tikzpicture}
    }
    \caption{An illustration of our MILP-based DFA framework.}
    \label{fig:framework_diagram}
\end{figure}

First, we automate and advance the search process for differential trails with a single solution using Mixed-Integer Linear Programming (MILP), significantly expanding the search scope. While Jana et al.~\cite{jana2024more} employed a meet-in-the-middle strategy by propagating input and output differences separately, this approach suffers from rapidly increasing complexity as the number of rounds increases. In contrast, we model the difference propagation of the entire cipher using MILP constraints and utilize a solver to explicitly verify the uniqueness of the solution for a given input-output difference pair by checking the number of feasible solutions. This modeling-based approach efficiently handles complex propagation behaviors, enabling the robust identification of unique differential trails even in deeper rounds, which were difficult to reach with previous matching techniques. Provided that the MILP model is correctly implemented, the fact that a single solution satisfies the constraints serves as a structural invariant that guarantees the trail is unique, regardless of the model's optimality or the solver's execution speed.

Second, we present a method to determine the optimal fault injection strategy by systematically analyzing all 128 possible single bit-flip faults. The efficiency of key recovery using a differential trail with a single solution is closely tied to the Linear Structure (LS) properties of the S-box. Specifically, an LS difference provides no key filtering capability because its corresponding entry in the Difference Distribution Table (DDT) is maximal (e.g., 16 for a 4-bit S-box), meaning it imposes no restrictions on the key candidates. Consequently, a trail containing a higher number of non-LS differences is essential for effective key filtering. However, in attacks on deeper rounds, an increase in non-LS differences paradoxically lowers the probability that a differential trail with a single solution exists, creating a trade-off. We address this by evaluating both the expected filtering capability and the probability of a unique trail's existence for every bit-flip position. This systematic analysis allows us to select the optimal fault that minimizes the total attack complexity.

Third, we provide a precise theoretical derivation of the required number of fault injections, supported by an analysis of the key space reduction behavior. We calculate the size of the remaining key candidate space to determine the exact number of faults required to recover the unique key. Crucially, this calculation explicitly accounts for the probability of a unique trail being found, thereby accurately reflecting the total expected number of injections, including any re-injections required when a fault fails to yield a differential trail with a single solution.

We applied this framework to \de{}, achieving results that surpass those of previous studies. For \de{} with the simple key schedule, we achieved a 6-round DFA with 3 faults, and 7- and 8-round DFAs with 2 faults each. For the rotating key schedule, the attacks require only $28+\alpha$, $13+\alpha$, and $11+\alpha$ faults for 6-, 7-, and 8-round DFAs, respectively. Furthermore, we applied our framework to \bak{}~\cite{baksi2023baksheesh}, successfully extending the attack range from the 3 rounds achieved by Jana et al.~\cite{jana2024more} to 5 rounds. However, since these results are less efficient compared to the recently proposed ToFA (Truncated impossible differential-based Fault Analysis), we omit the detailed description of the key recovery for \bak{} in this paper. A detailed comparison is provided in Table~\ref{tab:dfa-comparison}.

\begin{table}
\centering
\small
\setlength{\tabcolsep}{4pt}
\renewcommand{\arraystretch}{1.15}
\caption{Comparison of DFA attacks on \de{}.}\label{tab:dfa-comparison}
\begin{tabularx}{\textwidth}{X l c c c c}
\toprule
Cipher & Attack & Rnd & \#F & KS & Ref. \\
\midrule
\de{} (Simple)   & Information-Combining & 1      & 16          & $2^{39}$ & \cite{nageler2022information} \\
                 & SDFA & 1      & 64$\sim$128         & 1      & \cite{jana2024more} \\
                 & Unique Diff. Trail    & 2      & 64             & $2^{32}$ & \cite{jana2024more} \\
                 & Information-Combining & 3  & 16               & $2^{20}$ & \cite{nageler2022information} \\
                 
                 & Unique Diff. Trail    & 3      & 34             & $1$      & \cite{jana2024more} \\
                 & Unique Diff. Trail    & 4      & 16             & $1$      & \cite{jana2024more} \\
                 & Unique Diff. Trail    & 5      & 5               & $1$      & \cite{jana2024more} \\
                 
                 & MIFA & 6      & 3                   & 1      & Sec.~\ref{sec:improved_dfa_simple} \\
                 & MIFA & 7      & 2                   & 1      & Sec.~\ref{sec:improved_dfa_simple} \\
                 & MIFA & 8      & 2                   & 1      & Sec.~\ref{sec:improved_dfa_simple} \\
\midrule
\de{} (Rotating) & Information-Combining & 1   & $1728+\alpha$             & $1$      & \cite{nageler2022information} \\
                 & Information-Combining & 1      & $288+\alpha$               & $2^{32}$ & \cite{nageler2022information} \\
                 & SDFA & 1      & 64$\sim$128                   & 1      & \cite{jana2024more} \\
                 & Information-Combining & 3  & $(84\pm15)+\alpha$           & $1$      & \cite{nageler2022information} \\
                 & Unique Diff. Trail    & 3      & $96+\alpha$                 & $1$      & \cite{jana2024more} \\
                 & Unique Diff. Trail    & 4      & $48+\alpha$                 & $1$      & \cite{jana2024more} \\
                 & Unique Diff. Trail    & 5      & $36+\alpha$                   & $1$      & \cite{jana2024more} \\
                 & MIFA & 6     & $28+\alpha$                        & 1      & Sec.~\ref{sec:improved_dfa_rotating} \\
                 & MIFA & 7     & $13+\alpha$                        & 1      & Sec.~\ref{sec:improved_dfa_rotating} \\
                 & MIFA & 8     & $11 + \alpha$                        & 1      & Sec.~\ref{sec:improved_dfa_rotating} \\
\bottomrule
\end{tabularx}
\raggedright\footnotesize{
Rnd: Target rounds from the end of the cipher; \#F: Number of injected faults; KS: Remaining key space. \\
$\alpha$ signifies the number of faults injected into the \dc{}. \\
}
\end{table}

These attacks were validated through simulations. Our experiments confirmed that the probability of obtaining a differential trail with a single solution becomes negligible for attacks targeting 9 rounds or more on \de{}. This strongly suggests that our proposed attack has reached the maximal feasible depth for this approach, effectively establishing the practical upper bound for DFAs relying on unique trails against this cipher.

\subsection*{Organization} 

The remainder of this paper is organized as follows. Section~\ref{sec:preliminaries} introduces the necessary preliminaries, including an overview of DFA, the specifications of \de{}, the background on DFA based on differential trails with a single solution, and MILP-based methods for searching for differential trails. In Section~\ref{sec:framework}, we present our MILP-based framework for improving DFA. We apply the proposed framework to \de{} in Section~\ref{sec:application_default} to perform a key recovery attack and derive the precise fault complexity. Finally, we conclude the paper in Section~\ref{sec:conclusion}. The implementation code for our framework is publicly available online.\footnote{\url{https://github.com/shb115/MIFA}}

\section{Preliminaries} \label{sec:preliminaries}

\subsection{Differential Fault Attack}

Differential Fault Attack (DFA) is a cryptanalytic technique that recovers secret keys by injecting intentional faults into a cryptographic implementation and analyzing the differences between correct and faulty outputs. Specifically, DFA treats the fault injected into an intermediate state as a specific input difference $\Delta_{\text{in}}$. For instance, if an attacker induces a fault at the input of an S-box in the final round, they can obtain the correct ciphertext $c$ and the faulty ciphertext $c'$. The attacker can then verify a key candidate $k$ by checking whether the following equation holds:
$$
S^{-1}(c \oplus k) \oplus S^{-1}(c' \oplus k) = \Delta_{\textnormal{in}}.
$$
Key candidates that fail to satisfy this equality are eliminated. Specifically, when guessing a sub-key corresponding to the S-box size, the number of remaining candidates satisfying the equation is exactly equal to the value of the entry in the DDT corresponding to the input and output differences. Consequently, utilizing difference pairs associated with smaller values in the DDT yields stronger filtering power, thereby allowing for a more efficient reduction of the key search space. Since DFA typically targets the final operations of a cipher to extract the key, we establish the following convention: throughout this paper, any reference to targeting $n$ rounds (e.g., a 6-round or 7-round DFA) inherently denotes an attack on the last $n$ rounds of the cipher, counted from the end, unless explicitly stated otherwise.

\subsection{Descriptions of \de{} and \bak{}}

In this section, we provide the detailed specifications and structural properties of the two block ciphers analyzed in our study.

\subsubsection{Description of \de{}}

\de{} is a 128-bit block cipher proposed by Baksi et al. at ASIACRYPT 2021, designed to provide algorithm-level resistance against DFA. Its structure is given by  
$$
E = E_{\dl{}} \circ E_{\dc{}} \circ E_{\dl{}},
$$  
where the two $E_{\dl{}}$ parts consist of 28 rounds each and the $E_{\dc{}}$ part consists of 24 rounds, resulting in 80 rounds in total. \dl{} mainly provides DFA resistance, while \dc{} is designed to resist classical cryptanalysis.

\medskip\noindent\texttt{SubCells}. Each 4-bit nibble of the state is substituted by an S-box. \dl{} uses S-box $S$ with four LS elements, while \dc{} uses S-box $S_{core}$ with no non-trivial LS element, as detailed in Table~\ref{tab:S-boxes and their DDTs used in}.

\begin{figure*}[t]
  \centering
  \resizebox{1\linewidth}{!}{\input{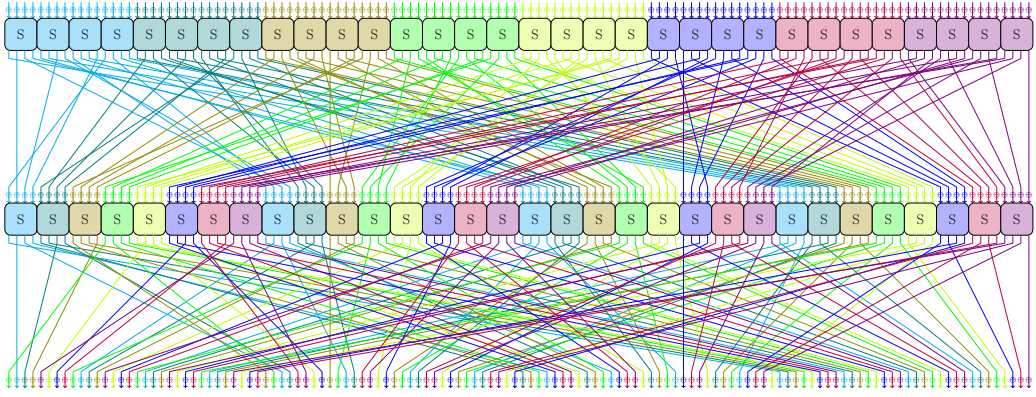}}
  \caption{QR group structure of \de{}} \label{fig:Two rounds of DEFAULT}
\end{figure*}

\medskip\noindent\texttt{PermBits}. A bit permutation adopted from GIFT-128~\cite{banik2017gift} is applied. It forms a Quotient-Remainder (QR) group structure, where the 16 bits coming from 4 S-boxes in one round affect only 4 S-boxes in the next round. The figure illustrating the QR group structure can be found in Figure~\ref{fig:Two rounds of DEFAULT}.

\medskip\noindent \texttt{AddRoundConstants}. A 6-bit round constant is XORed to specific bit positions (23, 19, 15, 11, 7, 3), and the most significant bit (bit 127) is toggled in every round to break symmetry between rounds.

\medskip\noindent \texttt{AddRoundKey}.
A 128-bit round key is XORed to the state. \de{} specifies two types of key schedules.

\medskip\noindent\textbf{Simple Key Schedule.} In this variant, the same 128-bit master key $K$ is applied as the round key for every round. That is, for any round $r$, $K_r = K$.

\medskip\noindent\textbf{Rotating Key Schedule.} This variant derives four distinct 128-bit round keys, denoted by $K_0, K_1, K_2,$ and $K_3$, from the master key. Specifically, $K_0$ is set to the master key, and each subsequent key $K_{i+1}$ is generated by applying 4 rounds of \dl{} operations to $K_i$. In each round $r$, the round key is determined cyclically as $K_{r \bmod 4}$.

\begin{table}[htbp]
  \centering
  \caption{S-boxes and their DDTs used in \de{}}
  \label{tab:S-boxes and their DDTs used in}
  \begin{subtable}[t]{0.48\linewidth}
    \centering
    \caption{\textsc{\dl{}} S-box $S$}
    \vspace{0.3em}
    \setlength{\tabcolsep}{1.2pt}
    \resizebox{\linewidth}{!}{%
    \begin{tabular}{@{}l *{16}{c}@{}}
      \toprule
      $u$      & \texttt{0} & \texttt{1} & \texttt{2} & \texttt{3} & \texttt{4} & \texttt{5} & \texttt{6} & \texttt{7} & \texttt{8} & \texttt{9} & \texttt{a} & \texttt{b} & \texttt{c} & \texttt{d} & \texttt{e} & \texttt{f} \\
      \midrule
      $S(u)$   & \texttt{0} & \texttt{3} & \texttt{7} & \texttt{e} & \texttt{d} & \texttt{4} & \texttt{a} & \texttt{9} & \texttt{c} & \texttt{f} & \texttt{1} & \texttt{8} & \texttt{b} & \texttt{2} & \texttt{6} & \texttt{5} \\
      \bottomrule
    \end{tabular}
    }
  \end{subtable}\hfill
  \begin{subtable}[t]{0.48\linewidth}
    \centering
    \caption{\dc{} S-box $S_{\text{core}}$}
    \vspace{0.3em}
    \setlength{\tabcolsep}{1.2pt}
    \resizebox{\linewidth}{!}{%
    \begin{tabular}{@{}l *{16}{c}@{}}
      \toprule
      $u$ & \texttt{0} & \texttt{1} & \texttt{2} & \texttt{3} & \texttt{4} & \texttt{5} & \texttt{6} & \texttt{7} & \texttt{8} & \texttt{9} & \texttt{a} & \texttt{b} & \texttt{c} & \texttt{d} & \texttt{e} & \texttt{f} \\
      \midrule
      $S_{\text{core}}(u)$ & \texttt{1} & \texttt{9} & \texttt{6} & \texttt{f} & \texttt{7} & \texttt{c} & \texttt{8} & \texttt{2} & \texttt{a} & \texttt{e} & \texttt{d} & \texttt{0} & \texttt{4} & \texttt{3} & \texttt{b} & \texttt{5} \\
      \bottomrule
    \end{tabular}
    }
  \end{subtable}

  \vspace{0.8em}

  \begin{subtable}[t]{0.48\linewidth}
    \centering
    \caption{DDT of $S$}
    \vspace{0.3em}
    \setlength{\tabcolsep}{1.2pt}
    \renewcommand{\z}{\(\cdot\)}
    \resizebox{\linewidth}{!}{%
    \begin{tabular}{@{}c *{16}{>{\centering\arraybackslash}p{0.75em}}@{}}
      \toprule
      $I\!\backslash\!O$ 
      & \texttt{0}&\texttt{1}&\texttt{2}&\texttt{3}&\texttt{4}&\texttt{5}&\texttt{6}&\texttt{7}&\texttt{8}&\texttt{9}&\texttt{a}&\texttt{b}&\texttt{c}&\texttt{d}&\texttt{e}&\texttt{f}\\
      \midrule
      \texttt{0} & \textcolor{red}{16} & \z & \z & \z & \z & \z & \z & \z & \z & \z & \z & \z & \z & \z & \z & \z \\
      \texttt{1} & \z  & \z & \z & 8 & \z  & \z & \z & \z & \z & 8  & \z & \z & \z & \z & \z & \z \\
      \texttt{2} & \z  & \z & \z & \z & \z  & \z & \z & 8  & \z & \z  & \z & \z & \z & 8  & \z & \z \\
      \texttt{3} & \z  & \z & \z & \z & 8  & \z & \z & \z & \z & \z  & \z & \z & \z & \z  & 8 & \z \\
      \texttt{4} & \z  & \z & \z & \z & \z  & \z & \z & 8  & \z & \z  & \z & \z & \z & 8  & \z & \z \\
      \texttt{5} & \z  & \z & \z & \z & 8  & \z & \z & \z & \z & \z  & \z & \z & \z & \z  & 8 & \z \\
      \texttt{6} & \z  & \z & \z & \z & \z  & \z & \z & \z & \z & \z  & \textcolor{red}{16}& \z & \z & \z  & \z & \z \\
      \texttt{7} & \z  & \z & \z & 8  & \z  & \z & \z & \z & \z & 8  & \z & \z & \z & \z  & \z & \z \\
      \texttt{8} & \z  & \z & \z & \z & \z  & \z & 8  & \z & \z & \z  & \z & \z & 8  & \z  & \z & \z \\
      \texttt{9} & \z  & \z & \z & \z & \z  & \z & \z & \z & \z & \z  & \z & \z & \z & \z  & \z & \textcolor{red}{16} \\
      \texttt{a} & \z  & 8  & \z & \z & \z  & \z & \z & \z & \z & \z  & \z & 8  & \z & \z  & \z & \z \\
      \texttt{b} & \z  & \z  & 8  & \z & \z  & \z & \z & \z & 8  & \z  & \z & \z & \z & \z  & \z & \z \\
      \texttt{c} & \z  & 8  & \z & \z & \z  & \z & \z & \z & \z & \z  & \z & 8  & \z & \z  & \z & \z \\
      \texttt{d} & \z  & \z  & 8  & \z & \z  & \z & \z & \z & 8  & \z  & \z & \z & \z & \z  & \z & \z \\
      \texttt{e} & \z  & \z  & \z  & \z & \z  & \z & 8  & \z & \z & \z  & \z & \z & 8  & \z  & \z & \z \\
      \texttt{f} & \z  & \z  & \z  & \z & \z  & \textcolor{red}{16}& \z & \z & \z & \z  & \z & \z & \z & \z  & \z & \z \\
      \bottomrule
    \end{tabular}
    }
  \end{subtable}\hfill
  \begin{subtable}[t]{0.48\linewidth}
    \centering
    \caption{DDT of $S_{\text{core}}$}
    \vspace{0.3em}
    \setlength{\tabcolsep}{1.2pt}
    \renewcommand{\z}{\(\cdot\)}
    \resizebox{\linewidth}{!}{%
    \begin{tabular}{@{}c *{16}{>{\centering\arraybackslash}p{0.75em}}@{}}
      \toprule
      $I\!\backslash\!O$ 
      & \texttt{0}&\texttt{1}&\texttt{2}&\texttt{3}&\texttt{4}&\texttt{5}&\texttt{6}&\texttt{7}&\texttt{8}&\texttt{9}&\texttt{a}&\texttt{b}&\texttt{c}&\texttt{d}&\texttt{e}&\texttt{f}\\
      \midrule
      \texttt{0} & 16& \z& \z& \z& \z& \z& \z& \z& \z& \z& \z& \z& \z& \z& \z& \z \\
      \texttt{1} & \z & \z& \z& \z& 2 & \z& \z& 2 & 2 & 2 & 2 & 2 & \z& 2 & 2 & \z \\
      \texttt{2} & \z & \z& \z& \z& \z& \z& 4 & 4 & \z& \z& \z& \z& \z& \z& 4 & 4 \\
      \texttt{3} & \z & 2 & \z& 2 & 2 & 2 & \z& \z& 2 & \z& 2 & \z& \z& \z& 2 & 2 \\
      \texttt{4} & \z & \z& \z& \z& \z& 4 & 4 & \z& \z& \z& \z& \z& \z& 4 & 4 & \z \\
      \texttt{5} & \z & \z& \z& \z& 2 & \z& \z& 2 & 2 & 2 & 2 & 2 & \z& 2 & 2 & \z \\
      \texttt{6} & \z & 4 & \z& 4 & \z& \z& \z& \z& \z& 4 & \z& 4 & \z& \z& \z& \z \\
      \texttt{7} & \z & 2 & \z& 2 & 2 & 2 & \z& \z& 2 & \z& 2 & \z& \z& \z& 2 & 2 \\
      \texttt{8} & \z & \z& \z& 4 & \z& \z& \z& 4 & \z& \z& \z& 4 & \z& \z& \z& 4 \\
      \texttt{9} & \z & \z& 2 & 2 & 2 & \z& 2 & \z& 2 & 2 & \z& \z& \z& 2 & \z& 2 \\
      \texttt{a} & \z & 4 & \z& \z& \z& \z& \z& \z& \z& 4 & \z& \z& 8 & \z& \z& \z \\
      \texttt{b} & \z & 2 & 2 & \z& 2 & 2 & 2 & 2 & 2 & \z& \z& 2 & \z& \z& \z& \z \\
      \texttt{c} & \z & \z& 4 & \z& \z& 4 & \z& \z& \z& \z& 4 & \z& \z& 4 & \z& \z \\
      \texttt{d} & \z & \z& 2 & 2 & 2 & \z& 2 & \z& 2 & 2 & \z& \z& \z& 2 & \z& 2 \\
      \texttt{e} & \z & \z& 4 & \z& \z& \z& \z& \z& \z& \z& 4 & \z& 8 & \z& \z& \z \\
      \texttt{f} & \z & 2 & 2 & \z& 2 & 2 & 2 & 2 & 2 & \z& \z& 2 & \z& \z& \z& \z \\
      \bottomrule
    \end{tabular}
    }
  \end{subtable}
\end{table}

\paragraph{Linear Structure and DFA Resistance.}
To understand the DFA resistance of \de{}, it is essential to define the Linear Structure.

\begin{definition}[Linear Structure]
For an S-box $S$, if a pair $(\Delta,\nabla)$ satisfies  
\[
S(x) \oplus S(x \oplus \Delta) = \nabla
\]  
for all inputs $x$, then $(\Delta,\nabla)$ is called a linear structure (LS) of $S$.
\end{definition}

For every S-box, (0,0) is always an LS. We therefore refer to any other LS as a non-trivial LS, and an S-box that has at least one non-trivial LS is called an LS S-box. The S-box used in \dl{} is a 4-bit S-box with four LS elements; when these appear as input and output differences, no key candidates can be eliminated. Moreover, \cite{baksi2022default} claimed that due to the properties of the LS, it is impossible to reduce the key space to one, regardless of the number of faults injected. For a cipher based on an LS S-box, assume a fault is injected into the input of an S-box one round before the last. The attacker knows the correct ciphertext $c$, the faulty ciphertext $c'$, and the injected input difference $\Delta_{\text{in}}$, but does not know the actual S-box input $u$ or the faulty input $u' = u \oplus \Delta_{\text{in}}$. If a pair $(u, k)$ exists such that $S(u) \oplus k = c$ and $S(u \oplus \Delta_{\text{in}}) \oplus k = c'$, then due to the LS property, the pair $(u \oplus \Delta, k \oplus \nabla)$ also produces the same ciphertext pair $(c, c')$. In other words, if a key candidate $k$ is possible, then $k \oplus \nabla$ is also possible for any LS output difference $\nabla$, meaning multiple key candidates will always remain. Therefore, the size of the key candidate space cannot be reduced below $(\#\text{LS})^{\#\text{S-boxes}}$. For example, in \de{}, this leaves $4^{32} = 2^{64}$ key candidates.

\subsubsection{Description of \bak{}}

\bak{}~\cite{baksi2023baksheesh} is a lightweight block cipher with a 128-bit block size and a 128-bit key, designed based on the GIFT-128 structure. It consists of 35 rounds, and each round is composed of the following four operations.

\texttt{SubCells}. A 4-bit S-box is applied to each 4-bit nibble of the state. The S-box contains a single non-trivial LS element, as detailed in Table~\ref{tab:S-box and DDT of}.

\begin{table}[htbp]
\centering
\caption{S-box and DDT of \bak{}}
\label{tab:S-box and DDT of}
\begin{subtable}{\textwidth}
\centering
\caption{The \bak{} S-box $S$}
\renewcommand{\arraystretch}{1.2}
\begin{tabular}{c|cccccccccccccccc}
\texttt{x} & \texttt{0} & \texttt{1} & \texttt{2} & \texttt{3} & \texttt{4} & \texttt{5} & \texttt{6} & \texttt{7} & \texttt{8} & \texttt{9} & \texttt{a} & \texttt{b} & \texttt{c} & \texttt{d} & \texttt{e} & \texttt{f} \\ \hline
\texttt{S(x)} & \texttt{3} & \texttt{0} & \texttt{6} & \texttt{d} & \texttt{b} & \texttt{5} & \texttt{8} & \texttt{e} & \texttt{c} & \texttt{f} & \texttt{9} & \texttt{2} & \texttt{4} & \texttt{a} & \texttt{7} & \texttt{1}
\end{tabular}
\end{subtable}

\begin{subtable}[t]{\textwidth}
\centering
\caption{DDT of $S$}
\renewcommand{\arraystretch}{1.0}
\setlength{\tabcolsep}{3pt}
\renewcommand{\z}{\(\cdot\)}
\begin{tabular}{@{}c *{16}{>{\centering\arraybackslash}p{1em}}@{}}
      \toprule
      $I\!\backslash\!O$ & \texttt{0} & \texttt{1} & \texttt{2} & \texttt{3} & \texttt{4} & \texttt{5} & \texttt{6} & \texttt{7} & \texttt{8} & \texttt{9} & \texttt{a} & \texttt{b} & \texttt{c} & \texttt{d} & \texttt{e} & \texttt{f} \\ \hline
\texttt{0} & \textcolor{red}{16} & \z & \z & \z & \z & \z & \z & \z & \z & \z & \z & \z & \z & \z & \z & \z \\
\texttt{1} &  \z & \z & \z & 4 & \z & \z & 4 & \z & \z & \z & \z & 4 & \z & \z & 4 & \z \\
\texttt{2} &  \z & \z & \z & 4 & \z & 4 & \z & \z & \z & \z & \z & 4 & \z & 4 & \z & \z \\
\texttt{3} &  \z & \z & \z & \z & \z & 4 & 4 & \z & \z & \z & \z & \z & \z & 4 & 4 & \z \\
\texttt{4} &  \z & \z & \z & 4 & \z & 4 & \z & \z & 4 & \z & \z & \z & \z & \z & 4 & \z \\
\texttt{5} &  \z & \z & \z & \z & \z & 4 & 4 & \z & 4 & \z & \z & 4 & \z & \z & \z & \z \\
\texttt{6} &  \z & \z & \z & \z & \z & \z & \z & \z & 4 & \z & \z & 4 & \z & 4 & 4 & \z \\
\texttt{7} &  \z & \z & \z & 4 & \z & \z & 4 & \z & 4 & \z & \z & \z & \z & 4 & \z & \z \\
\texttt{8} &  \z & \z & \z & \z & \z & \z & \z & \z & \z & \z & \z & \z & \z & \z & \z & \textcolor{red}{16}\\
\texttt{9} &  \z & 4 & \z & \z & 4 & \z & \z & \z & \z & 4 & \z & \z & 4 & \z & \z & \z \\
\texttt{a} &  \z & \z & 4 & \z & 4 & \z & \z & \z & \z & \z & 4 & \z & 4 & \z & \z & \z \\
\texttt{b} &  \z & 4 & 4 & \z & \z & \z & \z & \z & \z & 4 & 4 & \z & \z & \z & \z & \z \\
\texttt{c} &  \z & 4 & \z & \z & \z & \z & \z & 4 & \z & \z & 4 & \z & 4 & \z & \z & \z \\
\texttt{d} &  \z & \z & \z & \z & 4 & \z & \z & 4 & \z & 4 & 4 & \z & \z & \z & \z & \z \\
\texttt{e} &  \z & 4 & 4 & \z & 4 & \z & \z & 4 & \z & \z & \z & \z & \z & \z & \z & \z \\
\texttt{f} &  \z & \z & 4 & \z & \z & \z & \z & 4 & \z & 4 & \z & \z & 4 & \z & \z & \z \\

\bottomrule
\end{tabular}
\end{subtable}
\end{table}

\medskip\noindent
\texttt{PermBits}. The bits of the state are permuted using a bit permutation similar to that of GIFT-128 (QR group structure) but with different bit positions.

\medskip\noindent
\texttt{AddRoundConstants}. At each round, a 6-bit round constant and one additional toggle bit are XORed into specific bit positions of the state.

\medskip\noindent
\texttt{AddRoundKey}. The round key is XORed with the state. The first round key is the master key itself, and each subsequent round key is generated by rotating the previous key by one bit to the right.

\subsection{DFA Based on Differential Trails with a Single Solution}

Jana et al.~\cite{jana2024more} proposed an efficient DFA technique that exploits differential trails with a single solution. The attack begins by injecting a single bit-flip fault several rounds prior to the last to obtain a pair of correct and faulty ciphertexts. Using the known fault difference and the observed ciphertext difference, they employ a meet-in-the-middle strategy by propagating the differences forward and backward to deduce the exact intermediate differences. This process identifies a specific differential trail that guarantees a single, unique solution for the internal states.

This unique trail is then utilized for key recovery; however, since the LS of the S-box inherently leaves $2^{64}$ key candidates at the last round regardless of the number of injected faults, it prevents the immediate recovery of a unique key. To overcome this structural limitation, the attacker combines key guesses and performs partial decryption to the upper rounds. By verifying whether the difference of the decrypted intermediate state exactly matches the deterministic intermediate differences specified by the pre-computed unique trail, incorrect key candidates are effectively filtered out. This filtering process is repeated, iteratively reducing the key space until the unique key is successfully recovered.

\subsection{MILP-based Search for Differential Trails}

Mixed-Integer Linear Programming (MILP) is a mathematical optimization paradigm used to optimize an objective function subject to a set of linear constraints, where some variables are restricted to integer values. Since its introduction to cryptography by Mouha et al.~\cite{mouha2011differential} and Sun et al.~\cite{sun2014automatic}, it has been widely adopted for various cryptanalytic tasks, including finding differential trails~\cite{sun2014automatic} and identifying impossible differentials~\cite{sasaki2017new,cui2021new}. Furthermore, alongside MILP, SAT, and SMT solvers, recent works such as Hadipour et al.~\cite{hadipour2023finding,hadipour2024improved} have demonstrated that Constraint Programming (CP) is highly effective in automating the search for impossible differential, zero correlation, and integral attacks.

Following the standard modeling methodology established in~\cite{sun2014automatic}, where a differential trail represents the sequence of intermediate differences propagating through the cipher, the difference of each bit in the cipher state is represented by a binary variable. For non-linear operations like S-boxes, feasible input-output difference patterns are extracted from the DDT and converted into a set of linear inequalities. This conversion is typically achieved using techniques such as H-representation or logic minimization. Specifically, H-representation is a geometric approach that models the valid differential patterns as a convex hull defined by the intersection of half-spaces (linear inequalities). Alternatively, logic minimization tools like Espresso treat the valid patterns as the ON-set of a Boolean function, simplifying it to generate a minimal and compact set of linear constraints. Linear layers, including bit permutations and XOR operations, are modeled using equality constraints and XOR constraints, respectively. Operations involving constants or round keys are omitted as they do not affect differential propagation.

Based on this model, to find the optimal differential trail, the objective function is set to minimize the weight of the differential probability. This enables the solver to efficiently search for the trail with the maximum probability. Conversely, when analyzing impossible differentials, no objective function is defined. Instead, the target input difference $\Delta_{\text{in}}$ and output difference $\Delta_{\text{out}}$ are added as constraints to check the feasibility of a solution. If the solver returns an infeasible status, it indicates that the differential trail is structurally impossible.

\section{MIFA: MILP-based Differential Fault Attack Framework} \label{sec:framework}

In this section, we propose MIFA, a framework designed to enhance DFA through the use of an MILP solver. Since our framework and the subsequent analyses focus on the differential propagation within the core round functions, the methodologies presented here are independent of the specific key schedule and apply equally to both the Simple and Rotating Key Schedules of \de{}. First, we introduce an MILP-based method capable of identifying differential trails with a single solution that extend deeper into the cipher than those found by previous approaches. Attacks targeting deeper rounds are widely regarded as more potent, as they compel designers to implement countermeasures across more rounds. Moreover, longer unique differential trails allow for more frequent key filtering steps per fault, thereby enhancing filtering efficiency and reducing the total number of required fault injections. Furthermore, we present a method for determining the most effective bit positions for fault injection by systematically analyzing all possible bit-flip faults. This optimization further decreases the number of faults necessary for key recovery.

\subsection{MILP-based Search for Differential Trails with a Single Solution} \label{sec:MILP-based Search for Deterministic Trails}

In the preceding work by Jana et al.~\cite{jana2024more}, differential trails with a single solution were identified by propagating differences bidirectionally and verifying their collision at an intermediate round (meet-in-the-middle). However, as the number of rounds increases, this approach becomes computationally infeasible due to the exponential growth of the search space. To mitigate this limitation, we propose a new approach utilizing an MILP solver that efficiently searches for differential trails with a single solution even in deeper rounds.

The key contribution of our proposed method lies in a paradigm shift regarding the usage of the MILP solver. While traditional searches for impossible differentials focus on the absence of solutions (zero solutions), and standard differential characteristic searches focus on finding a trail with the optimal probability, our framework focuses on the uniqueness of the solution. Since the input and output differences obtained from actual physical fault injections are real phenomena, the MILP system modeling them must possess at least one feasible solution. If the solver returns exactly one solution for these constraints, it implies that the internal propagation trail is mathematically unique for the given input--output difference pair. This unique trail constitutes the differential trail with a single solution we seek.

The search process is summarized in the following steps:

\begin{description}
    \item[\textbf{Step 1.}] Inject a single bit-flip fault into the target round during encryption to obtain the correct ciphertext $C$ and the faulty ciphertext $C'$.    
    
    \item[\textbf{Step 2.}] Identify the input difference $\Delta_{\text{in}}$ corresponding to the fault injection position, compute the output difference $\Delta_{\text{out}} = C \oplus C'$, and set these values as constraints for the input and output variables in the MILP model.    
    
    \item[\textbf{Step 3.}] Model the differential propagation of the cipher following the standard framework proposed by Sun et al.~\cite{sun2014automatic}. 
    
    \item[\textbf{Step 4.}] Execute the solver to count the number of feasible solutions, denoted as $N_{\text{sol}}$.    
    
    \item[\textbf{Step 5.}] Proceed based on the value of $N_{\text{sol}}$:
    \begin{itemize}
        \item If $N_{\text{sol}} = 1$, the internal propagation trail is uniquely identified. Return the differential trail with a single solution and proceed to key recovery.
        \item If $N_{\text{sol}} > 1$, the unique trail cannot be determined. Return to \textbf{Step 1} to inject a new fault.
    \end{itemize}
\end{description}

We applied this framework to \de{} and \bak{}. Our experiments confirmed the feasibility of 6-, 7-, and 8-round DFAs on \de{} and 4- and 5-round DFAs on \bak{}. This was verified through 1,000 independent experiments for each cipher using random data. We injected a single bit-flip fault at position \texttt{0x1} for \de{} and \texttt{0x4} for \bak{}. The choice of the fault injection position significantly influences the attack complexity. We specifically selected these positions for this experiment because they exhibit the highest probability of yielding a differential trail with a single solution. A comprehensive analysis of all fault positions, which additionally takes key filtering efficiency into account, will be detailed in Section~\ref{sec:Searching for Effective Fault Injections for Key Recovery on DEFAULT}. The results are summarized in Table~\ref{tab:combined_results}.

\begin{table}[h]
\centering
\caption{Experimental results of finding unique differential trails for \de{} (with a \texttt{0x1} fault) and \bak{} (with a \texttt{0x4} fault). Here, ``\# Unique'' and ``\# Multiple'' denote the number of cases that resulted in a unique solution and multiple solutions, respectively.}
\label{tab:combined_results}
\begin{tabular}{c|c|c|c}
\toprule
Cipher & Fault on Round & \# Unique & \# Multiple \\
\midrule
\multirow{4}{*}{\de{}} & sixth-to-last   & 1000 & 0 \\
                       & seventh-to-last & 1000 & 0 \\
                       & eighth-to-last  & 999  & 1 \\
                       & ninth-to-last   & 0    & 12 \\
\midrule
\multirow{3}{*}{\bak{}} & fourth-to-last & 972 & 28 \\
                        & fifth-to-last  & 457 & 543 \\
                        & sixth-to-last  & 0   & 1000 \\ 
\bottomrule
\end{tabular}
\end{table}

The extremely low probability of finding unique solutions in deeper rounds (i.e., the ninth-to-last round for \de{} and the sixth-to-last round for \bak{}) is an inherent structural consequence. As the number of rounds increases, the cipher's diffusion causes the number of possible difference propagation branches to multiply exponentially. Consequently, it becomes highly improbable, though not strictly impossible, for a given pair of input and output differences to be satisfied by only a single unique trail.

All experiments were conducted on a workstation running Ubuntu 22.04 LTS with Gurobi~11.0.0, equipped with dual Intel Xeon Gold 6230R CPUs (providing a total of 104 threads) and 2~TB of RAM. Leveraging this multi-core environment, we utilized 100 parallel processes to execute the experiments. The runtime of the MILP solver generally increased with the number of analyzed rounds, although it varied depending on the specific input fault and output difference. For \de{}, trails for 6- and 7-round DFAs were obtained in under a second, and 8-round DFAs took about 10 seconds on average. However, as the round depth increases, the number of variables and the overall search space within the MILP model expand drastically due to the inherent exponential growth in search complexity. Consequently, the execution time for 9 rounds surged to approximately 24 hours per trail. This severe computational bottleneck restricted our feasible sample size, which explains why the 9-round evaluation was limited to 12 experiments rather than the 1,000 independent trials conducted for the 6-, 7-, and 8-round cases. For \bak{}, trails for 4- and 5-round DFAs were found in under a second, while the 6-round search averaged about 10 seconds.

\subsection{Searching for Effective Fault Injections for Key Recovery} \label{sec:Searching for Effective Fault Injections for Key Recovery on DEFAULT}

The efficiency of a DFA is heavily influenced by the bit position of the injected fault. Specifically, the S-boxes used in \dl{} and \bak{} possess a Linear Structure (LS), which creates a critical issue: for certain input differences (specifically \texttt{0x0}, \texttt{0x6}, \texttt{0x9}, or \texttt{0xf} for \dl{}, and \texttt{0x0} or \texttt{0x8} for \bak{}), the S-box produces a constant output difference regardless of the key value. This makes key filtering impossible. Consequently, selecting a fault that generates a high number of non-LS differences is essential for effective key filtering. Accurate estimation of the expected number of non-LS differences at each round is therefore a prerequisite for calculating the precise number of faults required for a successful attack.

We conducted a comprehensive search to identify the most effective single bit-flip faults. While there are 128 possible single bit-flip faults for a 128-bit state, theoretically analyzing all differential trails over the full number of rounds for every fault is computationally infeasible due to the exponential growth of the search space. To address this, we adopted a hybrid approach. We theoretically calculated the expected values for the initial rounds where exhaustive analysis is computationally manageable, and then switched to experimental simulations to estimate the probabilities for the subsequent deeper rounds.

In the theoretical analysis phase, we simulated the propagation of differences over the target rounds for all possible single bit-flip faults. Instead of exhaustively enumerating all individual differential trails, we tracked the probability distribution of differences based on the transition probabilities derived from the DDT of the S-box. Specifically, for a given input difference, we calculated the possible output differences and their associated probabilities, propagating this distribution through subsequent rounds. Based on this probabilistic propagation model, we calculated the total expected number of non-LS differences occurring across the analyzed rounds.

The analysis revealed that the expected value of non-LS differences varies significantly depending on the fault position, classifying them into distinct groups as shown in Table~\ref{tab:fault-efficiency-booktabs}. Note that in Table~\ref{tab:fault-efficiency-booktabs}, the fault classes for \texttt{0x2} and \texttt{0x4} in \de{} exhibit identical properties. This is because they form an equivalence class due to the structural symmetry and the \texttt{PermBits} layer of \de{}. For brevity, we discuss the \texttt{0x2} fault as a representative example, but the analysis applies equally to the \texttt{0x4} fault. 

\begin{table}[h!]
\centering
\captionsetup{skip=5pt}
\caption{Expected number of non-LS differences over three rounds for different faults in \de{} and \bak{}}
\label{tab:fault-efficiency-booktabs}
\setlength{\tabcolsep}{4pt}
\begin{tabular}{l l l}
\toprule
Cipher & Fault Pattern & Expected Non-LS differences \\ 
\midrule
\de{} & \texttt{0x2, 0x20, \ldots}, and \texttt{0x4, 0x40,\ldots}       & 26.75 (Highest) \\
& \texttt{0x8, 0x800, 0x80000, \ldots}           & 19.875 \\
& \texttt{0x80, 0x8000, 0x800000, \ldots}        & 19.625 \\
& \texttt{0x10, 0x1000, 0x100000, \ldots}       & 17.125 \\
& \texttt{0x1, 0x100, 0x10000, \ldots} & 16.875 (Lowest) \\ 
\midrule
\bak{} &\texttt{0x8, 0x800, 0x80000, \ldots}       & 31.375 (Highest) \\
&\texttt{0x80, 0x8000, 0x800000, \ldots}           & 31.125 \\
&\texttt{0x1, 0x10, \ldots}, and \texttt{0x2, 0x200, \ldots}        & 20.3125 \\
&\texttt{0x20, 0x2000, 0x200000, \ldots}       & 20.25 \\
&\texttt{0x40, 0x4000, 0x400000, \ldots} & 17.09375  \\ 
&\texttt{0x4, 0x400, 0x40000, \ldots} & 16.53125 (Lowest) \\ 
\bottomrule
\end{tabular}
\end{table}

Based on these expected values, we selected the optimal faults by balancing the high probability of generating non-LS differences (filtering efficiency) and the high probability of finding a differential trail with a single solution. A trade-off exists between these two factors: faults that generate many non-LS differences propagate rapidly, which benefits filtering but reduces the likelihood of a unique deterministic trail in deeper rounds. Instead of defining complex metrics, we adopted a single practical criterion: minimizing the total number of fault injections required for key recovery. We selected the fault that yields the lowest total fault count when considering both the success rate of finding a trail and the filtering power of that trail. Specifically, the total fault count defined in our metric is comprehensive; it accounts not only for the faults effectively used for key filtering but also for the expected number of wasted faults that fail to generate a differential trail with a single solution.

For \de{}, the \texttt{0x2} fault produces the most non-LS differences and is most efficient for filtering. However, while it guarantees a deterministic trail for 6- and 7-round attacks, it fails to do so consistently for 8-round attacks. In contrast, the \texttt{0x1} fault, despite being the least effective for filtering, maintains a 99.9\% probability of finding a deterministic trail even for an 8-round attack (as shown in Table~\ref{tab:combined_results}). Consequently, for deep-round attacks (e.g., 8 rounds), the \texttt{0x1} fault is the optimal choice despite its lower filtering efficiency per fault.

A similar trade-off is observed in \bak{}. The \texttt{0x8} fault provides the highest filtering efficiency and works well for 4-round attacks (88.8\% trail probability) but becomes impractical for 5-round attacks (0.9\%). Conversely, the \texttt{0x4} fault, which propagates more slowly, retains a viable 45.7\% probability for 5-round attacks. Thus, we selected the \texttt{0x4} fault as the optimal position for analyzing \bak{}.

For rounds beyond the reach of theoretical analysis, we estimated the average number of non-LS differences experimentally using 100,000 random trials. The empirical results obtained from these simulations are summarized in Table~\ref{tab:nib-wise-diff}. This estimation is crucial for determining the precise number of faults required. As the number of rounds increases, the input difference propagates through the cipher, causing the difference distribution to approach a uniformly random distribution. Accordingly, we utilize theoretical calculations for the initial rounds, switch to experimental estimation until the distribution converges to uniformity, and assume a uniform random distribution for all subsequent rounds. We apply this methodology in Section~\ref{sec:application_default} to calculate the probability of non-LS differences for our attack on \de{}.

\begin{table}[h]
\centering
\caption{Average difference occurrences per nibble in \de{} with \texttt{0x2} fault}
\label{tab:nib-wise-diff}
\setlength{\tabcolsep}{1.6pt}
\begin{tabular}{c|cccccccccccccccc}
\toprule
Round & \texttt{0} & \texttt{1} & \texttt{2} & \texttt{3} & \texttt{4} & \texttt{5} & \texttt{6} & \texttt{7} &
  \texttt{8} & \texttt{9} & \texttt{a} & \texttt{b} & \texttt{c} & \texttt{d} & \texttt{e} & \texttt{f} \\
\midrule
\multirow{1}{*}{$r-n+4$} 
& 7.27 & 3.85 & 2.55 & 1.60 & 4.48 & 2.32 & 1.12 & 0.68 
& 2.43 & 1.65 & 0.93 & 0.73 & 1.15 & 0.73 & 0.31 & 0.20 \\
\midrule
\multirow{1}{*}{$r-n+5$} 
& 2.76 & 2.64 & 2.06 & 1.90 & 2.65 & 2.57 & 1.95 & 1.84 
& 2.09 & 1.95 & 1.63 & 1.45 & 1.89 & 1.82 & 1.47 & 1.35 \\
\midrule
\multirow{1}{*}{$r-n+6$} 
& 2.03 & 2.03 & 1.99 & 2.00 & 2.04 & 2.03 & 2.00 & 2.00 
& 1.99 & 2.00 & 1.97 & 1.97 & 2.00 & 2.00 & 1.97 & 1.98 \\
\bottomrule
\end{tabular}
\raggedright\footnotesize{
$r$ denotes the round of the cipher, and $r-n$ denotes the fault injection round.
}
\end{table}

\section{Key Recovery Attack on DEFAULT and Analysis of Required Faults} \label{sec:application_default}

In this section, we apply the proposed MIFA framework to \de{} to perform a practical key recovery attack. Before detailing the attack procedure, we theoretically address the potential concern regarding irregular reductions in the number of candidate keys during filtering. We show that the number of candidate keys in nibble-level filtering always maintains a power-of-two form ($2^k$). Furthermore, we demonstrate that the lower bound of the candidate key space for group-level filtering is $2^4$. We then verify this property through experimental results and present the specific attack procedure along with the number of faults required for key recovery.

\subsection{Structural Properties of Key Space Reduction} \label{sec:structural_properties}

The core of DFA involves identifying candidate keys that satisfy the differential equation $S(u) \oplus S(u \oplus \Delta_{\text{in}}) = \Delta_{\text{out}}$. Through an in-depth analysis specifically targeting \de{}, we have identified novel structural properties regarding the key recovery process: the discrete reduction pattern ($2^k$) at the nibble level and the structural lower bounds at the group level determined by the properties of equivalent keys.

\paragraph{Nibble Level: Equivalent Keys and $2^k$ Reduction.}
Due to the LS properties of the S-box, valid key candidates form sets of indistinguishable keys, referred to as equivalent keys~\cite{nageler2022information}. Since the S-box of \de{} possesses 4 LS, each nibble contains 4 equivalent keys. We define this set of 4 keys as a single key group. Consequently, the initial 16 candidate keys consist of exactly 4 key groups. The DDT of the \de{} S-box comprises solely the values 0, 8, and 16. In the context of key groups, a DDT value of 8 indicates that exactly 2 key groups are possible. Since the elements of a key group are always preserved or eliminated together, the size of the candidate pool reduces from 4 key groups (16 keys) to 2 key groups (8 keys), and finally to 1 key group (4 keys), thereby always maintaining a power-of-two form ($2^k$).

\paragraph{Group Level: Lower Bound of $2^4$ Due to Equivalent Keys.}
At the group level, the lower bound of the key space is determined by the expansion of equivalent keys. In the structure of \de{}, a single group key affects 4 nibbles in the upper round, which are used for filtering. Each of these 4 nibbles independently possesses 4 equivalent keys. When combined at the group level, these result in a total of 16 indistinguishable keys. Consequently, regardless of the number of faults injected, the candidate keys within a group remain indistinguishable from one another. This makes it structurally impossible for the number of candidate keys at the group level to fall below $2^4$.

\paragraph{Half Level: Convergence to a Unique Key via Inter-Round Dependency.}
At the half level, a total of 16 nibbles are involved in the filtering process. Unlike the group level which spans two rounds, the half level extends the key guessing process over three rounds, applying the final filtering at this third round to the candidates that have already passed the two-round constraints. The equivalent key property, maintained at the nibble and two-round group levels, is no longer preserved because the diffusion layer thoroughly intertwines these 16 nibbles across this extended three-round span, inherently breaking local equivalences. Furthermore, as detailed in Appendix~\ref{app:Experimental Results}, our experimental results confirm this behavior, demonstrating that while the candidate space is bounded at $2^4$ at the group level, the number of candidate keys successfully converges to a unique key ($2^0$) upon reaching the half level.

These theoretical findings regarding the number of remaining candidate keys at each level were verified experimentally, as detailed in Appendix~\ref{sec:Experimental_Verification_of_Candidate_Key_Reduction}.

\subsection{Effective Fault Selection for \de{}}
\label{sec:effective_fault_selection}

Based on the analysis in Section~\ref{sec:Searching for Effective Fault Injections for Key Recovery on DEFAULT}, we select the input faults to minimize the total number of faults required for key recovery. Accordingly, we utilize \texttt{0x2} for the 6- and 7-round attacks, and \texttt{0x1} for the 8-round attack. The distribution of non-LS differences for these faults directly influences the calculation of the required number of fault injections presented in Section~\ref{sec:improved_dfa_simple}. 

Specifically, for the \texttt{0x2} fault, theoretical derivation is possible up to round $r-n+3$. The average number of non-LS differences propagates theoretically as 1, 3, 7.5, and 16.25 for rounds $r-n$ to $r-n+3$, respectively. For round $r-n+4$, experimental verification yields 21.76, and from round $r-n+5$ onward, it converges to approximately 24, which represents a uniform distribution.

Similarly, for the \texttt{0x1} fault, the sparsity of active trails permits theoretical derivation up to round $r-n+4$. The theoretical propagation sequence from round $r-n$ to $r-n+4$ is 1, 2, 4.5, 10.375, and 19.273, respectively. Subsequent experimental verification shows 23.4 for round $r-n+5$, and from round $r-n+6$ onward, it also converges to approximately 24.

\subsection{Improved DFA on \de{} with Simple Key Schedule}
\label{sec:improved_dfa_simple}

In \de{}, the simple key schedule applies the same master key to all rounds. Jana et al.~\cite{jana2024more} proposed a key filtering method that guesses the partial key to partially decrypt the correct and faulty ciphertexts one round at a time, and then compares the resulting difference against that of a differential trail with a single solution. The detailed attack procedure for \de{} with a simple key schedule consists of the following four main steps:

\begin{description}
    \item[Step 1: Filtering at the first-to-last round.] 
    The last round key is guessed nibble by nibble to perform a one-round partial decryption. The resulting differences are compared with the expected differences of the deterministic trail at the first-to-last round. If they do not match, the guessed key is discarded.

    \item[Step 2: Filtering at the second-to-last round.] 
    The filtered last round key is guessed group-wise, where a group consists of 4 nibbles at positions $\{0, 8, 16, 24\}$, $\{1, 9, 17, 25\}$, and so on. Then, the key at the first-to-last round is guessed nibble-wise. Using both guessed keys, partial decryption is performed to compare the resulting differences with the expected differences at the second-to-last round. At this stage, due to the simple key schedule, 4 nibbles at the first-to-last round are determined by the group key guessed in the last round (group-determined nibbles), while the others are not (non-group-determined nibbles). In the case of a non-group-determined nibble, the currently guessed group key is discarded only if all key candidates for that specific nibble, which survived Step 1, fail to satisfy the difference condition.

    \item[Step 3: Filtering at the third-to-last round.] 
    The filtered last round key is guessed half-wise, where a half consists of 4 groups (totaling 16 nibbles) corresponding to groups $\{0, 2, 4, 6\}$ and $\{1, 3, 5, 7\}$. Then, the key at the first-to-last round is guessed group-wise, and the key at the second-to-last round is guessed nibble-wise. Using these guessed keys, partial decryption is performed to compare the resulting differences with the expected differences at the third-to-last round. 
    
    At this stage, due to the simple key schedule, the 32 nibbles are classified into four distinct categories, each containing 8 nibbles: 
    (1) \textit{Fully-Determined}, where both the group key at the first-to-last round and the nibble key at the second-to-last round are determined by the guessed half key; 
    (2) \textit{Group-Determined}, where only the group key is determined; 
    (3) \textit{Nibble-Determined}, where only the nibble key is determined; and 
    (4) \textit{Non-Determined}, where neither key is determined. 
    
    For the cases where the group key, the nibble key, or both are not determined, the currently guessed half key is discarded if all corresponding candidate keys surviving from Steps 1 and 2 fail to satisfy the difference condition.
    
    \item[Step 4: Filtering before the third-to-last rounds.] 
    The filtered last round key is guessed as a full 128-bit key, and partial decryption is performed up to the target round. The key is discarded if the resulting differences do not match those of the corresponding round in the deterministic trail.
\end{description}

We calculate the number of faults required to reduce the key candidates to a unique value in a DFA on \de{} with a simple key schedule. We achieve this by tracking the size of the remaining key candidate space round by round, utilizing the identified differential trails with a single solution. Our calculation is based on the probabilities of non-LS differences occurring, as analyzed in Section~\ref{sec:effective_fault_selection}.

\subsubsection{6-Round DFA on \de{} with Simple Key Schedule}

We analyze the DFA on \de{} with a simple key schedule, where faults are injected at the sixth-to-last round. Let $n_t$ be the number of differential trails with a single solution.

\paragraph{Filtering at the first-to-last round.}
The reduction of the key candidate space begins at an initial size of $2^4$. Since the DDT entry value for a non-LS difference is 8, the occurrence of a single non-LS difference unconditionally reduces the candidate space to $2^3$. The probability that a specific nibble exhibits a non-LS difference is $24/32 = 3/4$. Specifically, the $2^4$ key space consists of 4 key groups, and a non-LS difference restricts candidates to 2 groups. Thus, when multiple non-LS differences occur, the probability that the surviving key groups completely overlap, resulting in a candidate size of $2^3$, is $(1/2)^n$. Therefore, the expected size of the key candidate space, given $n_t=3$, is calculated as:
$$
\left(\frac{1}{4}\right)^3 \cdot 2^4 + \left[ 3\left(\frac{1}{4}\right)^2\left(\frac{3}{4}\right) + 3\left(\frac{1}{4}\right)\left(\frac{3}{4}\right)^2 \left(\frac{1}{2}\right)^2 + \left(\frac{3}{4}\right)^3 \left(\frac{1}{2}\right)^3 \right] \cdot 2^3 + \frac{351}{512} \cdot 2^2 \approx 2^{2.43}
$$

\paragraph{Filtering at the second-to-last round.}
Assuming $n_t=3$, the initial candidate space for a key group is $(2^{2.43})^4 = 2^{9.72}$. Guessing a single group key in the last round enables key filtering for four nibbles in the second-to-last round. Since the DDT entry for a non-LS difference is 8, the probability that a specific key candidate satisfies the differential condition is $1/2$. This corresponds to the baseline of 1-bit filtering.

The number of group-determined nibbles is 4, and the total number of groups is 8. Considering the probability of a nibble exhibiting a non-LS difference, which is $21.76/32$, the average filtering bits per group is calculated as:
$$
\frac{4 \times (21.76/32)}{8} \approx 0.34
$$
bits.

For the remaining 28 non-group-determined nibbles, the currently guessed last round group key is discarded only if all nibble key candidates at the first-to-last round fail to satisfy the difference condition.
When the nibble key candidate space is $2^2$ (probability $351/512$), exactly one key group remains. Thus, the elimination probability is $1/2$, yielding full 1-bit filtering effectiveness:
$$
\frac{28 \times (21.76/32) \times (351/512)}{8} \approx 1.63
$$
bits.
When the candidate space is $2^3$ (probability $153/512$), two key groups remain. For the group key to be filtered, both remaining groups must be invalid. The probability of this simultaneous elimination is $(1/2) \times (1/2) = 1/4$. Since this is half the baseline elimination probability ($1/2$), we apply a weighting factor of $1/2$ to the filtering:
$$
\frac{28 \times (21.76/32) \times (153/512) \times (1/2)}{8} \approx 0.36
$$
bits.
Finally, when the candidate space is $2^4$, four key groups exist. Since it is impossible for all four groups to fail simultaneously, no filtering occurs in this case.

Consequently, a total average of $2.33$ bits are filtered per group key guess. Considering the structural lower bound at the group level, the number of remaining candidate keys after using $n_t$ trails is given by $\max(2^4, 2^{9.72-2.33n_t})$.

\paragraph{Filtering at the third-to-last round.}
Since a half consists of four groups, considering the structural lower bound of each group ($2^4$), the initial candidate space for a half-key at the last round is $\max(2^{16}, 2^{38.88-9.32n_t})$. A single half-key guess allows for filtering across 16 nibbles at the third-to-last round. We apply the same logic as in the previous round, using the non-LS difference probability of $16.25/32$ as calculated in the previous section.

The 32 nibbles involved are categorized based on their dependency on the half-key. For the eight \textit{fully-determined} nibbles, both the group key and the nibble key are fixed. Since the key is unique, the elimination probability is $1/2$. Thus, the average filtering per half is:
$$
\frac{8 \times (16.25/32)}{2} \approx 2.03
$$
bits.

For the eight \textit{group-determined} nibbles, the group key is determined, but the nibble key depends on the candidate keys surviving from previous steps. The currently guessed half-key is discarded only if all nibble key candidates fail to satisfy the difference condition. Specifically, when the nibble key candidate space is $2^2$ (probability $351/512$), exactly one key group remains, resulting in an elimination probability of $1/2$ and contributing approximately $1.39$ bits ($= \frac{8 \times (16.25/32) \times (351/512)}{2}$). When the candidate space is $2^3$ (probability $153/512$), two key groups remain. Since both key groups must fail simultaneously for the half-key to be filtered (probability $1/4$), we apply a weighting factor of $1/2$, resulting in a filtering effect of approximately $0.30$ bits ($= \frac{8 \times (16.25/32) \times (153/512) \times (1/2)}{2}$). In the case where the candidate space is $2^4$, four key groups remain; since it is practically impossible for all four to fail simultaneously, this case contributes no filtering.

Finally, for the remaining 16 nibbles (classified as nibble-determined or non-determined), the group key is not determined by the half-key guess, effectively making the candidate space $2^{16}$. Since the probability that all candidates in such a large space fail to satisfy the difference condition is negligible, we exclude this potential filtering from our complexity calculation.

Thus, a total average of $3.72$ ($=2.03+1.39+0.30$) bits are filtered per half-key guess. The final remaining key space is $2^{38.88 - 9.32n_t - 3.72n_t} = 2^{38.88-13.04n_t}$. As discussed in Section~\ref{sec:structural_properties}, upon reaching the half level, the key candidates can structurally converge to a unique solution ($2^0$).

\paragraph{Filtering at the preceding rounds.}
For filtering at the fourth-to-last round, the full key is guessed by combining the two halves. At this stage, we observe an average of 7.5 active S-boxes with non-LS differences. Since each non-LS difference provides 1 bit of filtering as established previously, the candidate space reduces to $(2^{38.88 - 13.04 n_t})^2 \cdot 2^{- 7.5 n_t} = 2^{77.76 - 33.58 n_t}$.
Considering the additional 3 and 1 non-LS differences occurring at the fifth- and sixth-to-last rounds respectively, the final number of remaining key candidates is:
$$
2^{77.76 - 33.58 n_t - 3 n_t - 1 n_t} = 2^{77.76 - 37.58 n_t}
$$
For $n_t=2$, the remaining candidate space is approximately $2^{2.6}$, which implies that while the candidates are reduced to a very small number (approx. 6), it is insufficient for unique key recovery. However, using $n_t=3$ reduces the space to $2^{77.76 - 37.58 \times 3} = 2^{-34.98}$. As this value is significantly less than 1, only the correct key remains. Therefore, injecting three faults is sufficient to recover the entire key, a result we have also verified experimentally.

\subsubsection{7-Round DFA on \de{} with Simple Key Schedule}

The complexity of the 7-round DFA is calculated in the same manner as the 6-round DFA.

\paragraph{Filtering at the first-to-last round.}
For $n_t=2$, the average number of remaining key candidates per nibble is calculated as:
$$
\frac{1}{16} \cdot 2^4 + \left[ 2 \left(\frac{1}{4}\right) \left(\frac{3}{4}\right) + \left(\frac{3}{4}\right)^2 \left(\frac{1}{2}\right)^2 \right] \cdot 2^3 + \left( 1 - \frac{1}{16} - \frac{33}{64} \right) \cdot 2^2 \approx 2^{2.77}
$$

\paragraph{Filtering at the second-to-last round.}
For the four group-determined nibbles, an average of $4 \times (24/32) / 8 = 0.375$ bits are filtered per group. For the 28 non-group-determined nibbles, the average filtering is:
$$
\frac{28 \times (24/32) \times (33/64)}{8} + \frac{28 \times (24/32) \times (27/64) \times (1/2)}{8} \approx 1.91
$$
bits per group. Thus, a total of $2.285$ ($\approx 0.375+1.91$) bits are filtered, reducing the candidate space to $\max(2^4,2^{11.08-2.285n_t})$.

\paragraph{Filtering at the third-to-last round.}
With a non-LS difference probability of $21.76/32$, the initial candidate space per half is $2^{44.32-9.14n_t}$. For the eight fully-determined nibbles, the average filtering is:
$$
\frac{8 \times (21.76/32)}{2} \approx 2.72
$$
bits per half. For the eight group-determined nibbles, the average filtering is:
$$
\frac{8 \times (21.76/32) \times (33/64)}{2} + \frac{8 \times (21.76/32) \times (27/64) \times (1/2)}{2} \approx 1.98
$$
bits per half. Thus, a total of $4.7$ ($=2.72+1.98$) bits are filtered, resulting in $2^{44.32-13.84n_t}$ candidates.

\paragraph{Filtering at the preceding rounds.}
The final key candidate space reduces to 
$$
2^{88.64-13.84 \cdot 2n_t - 27.75n_t} = 2^{88.64-55.43n_t}.
$$ 
Substituting $n_t=2$ yields $2^{88.64 - 110.86} = 2^{-22.22}$. Since this value is less than 1, injecting two faults is sufficient to recover the entire key.

\subsubsection{8-Round DFA on \de{} with Simple Key Schedule}

The complexity of the 8-round DFA is calculated similarly, but using the \texttt{0x1} input difference.

\paragraph{Filtering at the first-to-last round.}
For $n_t=2$, the average number of remaining key candidates per nibble is approximately $2^{2.77}$.

\paragraph{Filtering at the second-to-last round.}
The remaining key candidate space per group is $2^{11.08-2.285n_t}$.

\paragraph{Filtering at the third-to-last round.}
For the eight fully-determined nibbles, the average filtering is:
$$
\frac{8 \times (23.4/32)}{2} \approx 2.925
$$
bits per half. For the eight group-determined nibbles, the average filtering is:
$$
\frac{8 \times (23.4/32) \times (33/64)}{2} + \frac{8 \times (23.4/32) \times (27/64) \times (1/2)}{2} \approx 2.125
$$
bits. Thus, a total of $5.05$ ($\approx 2.925+2.125$) bits are filtered, reducing the candidate space to $2^{44.32-14.19n_t}$.

\paragraph{Filtering at the preceding rounds.}
The final key candidate space is reduced to 
$$
2^{88.64-14.19 \cdot 2n_t - 37.148n_t} = 2^{88.64-65.528n_t}.
$$
Substituting $n_t=2$ yields $2^{88.64 - 131.056} = 2^{-42.416}$. Since this value is less than 1, injecting two faults is sufficient to recover the entire key.

To validate these theoretical complexity estimates, we conducted a large-scale simulation. The detailed experimental results regarding the average number of remaining candidate keys at each filtering step are provided in Appendix~\ref{sec:Experimental_Verification_of_Candidate_Key_Reduction}.

\subsection{Improved DFA on \de{} with Rotating Key Schedule}
\label{sec:improved_dfa_rotating}

In \de{}, the rotating key schedule uses four independent round keys repeatedly. Nageler et al.~\cite{nageler2022information} demonstrated that due to this key schedule and the properties of the LS S-box, if $(k_0, k_1)$ is the correct key, an equivalent key pair $(\hat{k_0}, \hat{k_1})$ capable of performing the same decryption also exists. Consequently, the S-box of the \dl{} has 4 equivalent keys, the round function implies $2^{64}$ equivalent keys, and the entire \dl{} with the rotating key schedule possesses $(2^{64})^3=2^{192}$ equivalent keys.

The attack on \de{} with a rotating key schedule, as presented by Jana et al.~\cite{jana2024more}, consists of the following three steps:

\begin{description}
    \item[Step 1: Reduction to Equivalent Key Space.] This step reduces the key candidate space over four rounds, corresponding to the four independent keys used in the rotating key schedule. For the first three keys encountered in the reverse direction (e.g., $\hat{k}_3, \hat{k}_2, \hat{k}_1$), we utilize differential trails with a single solution to filter the key space, followed by reducing it to a normalized representative $\bar{k}_i$. For instance, if the candidate set is \{\texttt{0x0}, \texttt{0x5}, \texttt{0xa}, \texttt{0xf}\}, the normalized representative is $\bar{k}_i = \texttt{0x0}$. Partial decryption is performed using this $\bar{k}_i$ to proceed to the next round. For the $\hat{k}_0$, the candidate space is filtered to $2^{64}$ but not normalized. Consequently, the combination of the normalized keys $\bar{k}_1, \bar{k}_2, \bar{k}_3$ and the $2^{64}$ candidates for $\hat{k}_0$ forms the equivalent key space.

    \item[Step 2: Identification of the Unique Key.] In this step, faults are injected at a round corresponding to twice the number of rounds in Step 1. Partial decryption is performed using the normalized keys $\bar{k}_i$ obtained in Step 1 to identify a differential trail with a single solution. Using this trail, the candidate space for $\hat{k}_0$ is filtered from $2^{64}$ down to a unique value. Consequently, the multiple equivalent key spaces collapse into a single unique key space.

    \item[Step 3: Recovery of the Original Key.] Finally, faults are injected into the \dc{} to recover the master key from the identified equivalent key.
\end{description}

We apply the same optimization approach used for the simple key schedule to this rotating key schedule scenario to minimize the number of required faults. Specifically, by utilizing an MILP solver to identify differential trails with a single solution for 6-, 7-, and 8-round DFAs, we significantly reduce the fault requirements for Steps 1 and 2.

For Step 1, due to the diffusion properties of the cipher, a deterministic trail tends to exhibit more non-LS differences as it propagates through more rounds. This implies that if the key space can be reduced to $2^{64}$ with high probability at the fourth-to-last round, the same reduction can be achieved with even higher probability at subsequent rounds (first- to third-to-last). Therefore, our analysis focuses on calculating the required number of fault injections to achieve reduction at the fourth-to-last round.

For Step 2, identifying the key within the equivalent key space requires injecting additional faults at significantly earlier rounds (i.e., twelfth-, fourteenth-, and sixteenth-to-last rounds for the 6-, 7-, and 8-round attack scenarios, respectively). Attacking such deep rounds presents a problem analogous to the simple key schedule attack. Consequently, it is sufficient to inject only 3, 2, and 2 additional faults for each respective scenario to uniquely determine the equivalent key.

\subsubsection{6-Round DFA on \de{} with Rotating Key Schedule}

For a fault injected at position \texttt{0x2} at the sixth-to-last round, the probability of obtaining a non-LS difference at the fourth-to-last round is $7.5/32$. The probability that the key space fails to reduce to $2^2$ (which occurs when non-LS differences appear at most once, or appear multiple times but yield redundant information) can be modeled using the binomial distribution as:
$$
\left(\frac{24.5}{32}\right)^{n_t} + n_t \cdot \left(\frac{7.5}{32}\right) \cdot \left(\frac{24.5}{32}\right)^{n_t-1} + \sum_{i=2}^{n_t} \binom{n_t}{i} \left(\frac{7.5}{32}\right)^i \cdot \left(\frac{24.5}{32}\right)^{n_t-i}\cdot\left(\frac{1}{2}\right)^i
$$
We determined the minimum number of trails, $n_t$, such that this failure probability is less than 5\%, thereby guaranteeing a success rate of over 95\%. It is important to note that the final fault count derived from this $n_t$ represents the total number of random fault injections required to obtain these effective trails with high probability. Our analysis confirms that starting from $n_t=25$, the key space is reduced to $2^2$ with a probability exceeding 95\%. Including the three faults required for Step 2, a total of 28 faults are sufficient to recover the key in the 6-round attack scenario.

\subsubsection{7-Round DFA on \de{} with Rotating Key Schedule}

For a fault injected at position \texttt{0x2} at the seventh-to-last round, the probability of a non-LS difference appearing at the fourth-to-last round is $16.25/32$. Similar to the 6-round attack, we confirmed that starting from $n_t=11$, the key space of all nibbles reduces to $2^2$ with a probability greater than 95\%. Including the two faults required for Step 2, a total of 13 faults are sufficient to recover the equivalent key in the 7-round attack scenario.

\subsubsection{8-Round DFA on \de{} with Rotating Key Schedule}

For a fault injected at position \texttt{0x1} at the eighth-to-last round, the probability of a non-LS difference appearing at the fourth-to-last round is $19.27/32$. Similar to the 6- and 7-round attacks, we confirmed that starting from $n_t=9$, the key space of all nibbles reduces to $2^2$ with a probability exceeding 95\%. Including the two faults required for Step 2, a total of 11 faults are sufficient to recover the equivalent key in the 8-round attack scenario.

\section{Discussion and Conclusion}
\label{sec:conclusion}

Before concluding, we discuss the scalability and general applicability of our proposed framework. Regarding the fault model, our primary analysis relies on the single bit-flip assumption. We evaluated the framework under broader physical assumptions, including nibble, byte, and random multi-bit faults. However, empirical results indicated that such faults cause differences to propagate with significantly higher diffusion. This rapid propagation yields an unmanageable number of feasible solutions in the MILP model, making it practically infeasible to obtain a differential trail with a single solution for deeper rounds. Therefore, we conclude that the single bit-flip fault is the optimal injection model for DFA methodologies that exploit differential trails with a single solution.

Furthermore, regarding the applicability to other structural designs, our framework is generically applicable to any S-box. Nevertheless, a fundamental trade-off exists when targeting non-LS S-boxes. Ciphers utilizing non-LS S-boxes inherently exhibit superior differential properties, which trigger stronger key filtering power but drastically reduce the probability that a differential trail with a single solution exists over multiple rounds. Consequently, our framework based on differential trails with a single solution is less suitable for ciphers employing non-LS S-boxes, as finding such trails becomes inherently difficult.

In this paper, we proposed MIFA, a new framework that utilizes an MILP solver to significantly enhance the efficiency of DFA attacks. By systematically identifying differential trails with a single solution in deeper rounds, our approach enables successful attacks on rounds that were previously considered difficult to reach. As the number of attackable rounds increases, the efficiency of key filtering improves substantially, leading to a natural reduction in the required number of faults. We also emphasized that targeting these deeper rounds is of significant practical value, as countermeasures are often concentrated on the last few rounds.

Furthermore, we investigated the key filtering efficiency for all possible bit-flip faults. We selected the optimal bit position by comprehensively balancing the trade-off between key filtering efficiency and the probability of finding a differential trail with a single solution. A significant contribution of our work is the precise theoretical calculation of the required number of faults, achieved by proving that the key candidate space for LS S-box-based ciphers always reduces in the form of $2^k$ and by identifying the structural lower bound of key candidates during filtering at the second-to-last round.

Consequently, we achieved state-of-the-art results on DEFAULT. For DEFAULT with the simple key schedule, we demonstrated that 6-, 7-, and 8-round DFAs can be performed with only 3, 2, and 2 faults, respectively. For the rotating key schedule, our attacks require totals of 28, 13, and 11 faults for the same respective rounds. These results represent the optimal efficiency achievable among methods relying on differential trails with a single solution. Furthermore, we successfully applied our framework to BAKSHEESH, extending the attack range to 5 rounds; however, since these results are less efficient compared to the recently proposed ToFA~\cite{kundu2025tofa}, we omitted the detailed key recovery procedures from this paper.

\section*{Acknowledgements}

We would like to express our sincere gratitude to the anonymous reviewers and the shepherd for their insightful comments and constructive feedback, which significantly improved the quality of this manuscript. This research was supported in part by the Basic Science Research Program through the National Research Foundation of Korea (NRF) funded by the Ministry of Education (No. 25411243) and in part by the 2025 Research Grant from Kangwon National University.

\bibliographystyle{alpha}
\bibliography{biblio}

\appendix
\section{Experimental Verification of Candidate Key Reduction}
\label{sec:Experimental_Verification_of_Candidate_Key_Reduction}

To verify the theoretical key recovery complexity presented in Section~\ref{sec:improved_dfa_simple}, we conducted a large-scale simulation based on the 8-round attack scenario for \de{} with a simple key schedule. The goal of this experiment is to measure the average number of remaining candidate keys at the nibble, group, half, and state levels, which are updated at each filtering step.

\subsection{Experimental Setup and Data Selection}
We generated 1,010 random master keys and performed attacks by injecting 4 faults for each key. Although our theoretical analysis indicates that key recovery is feasible with fewer faults, we employed 4 faults in this experiment solely to expedite the experiment.

As discussed in Section~\ref{sec:MILP-based Search for Deterministic Trails}, randomly injected faults do not always yield a differential trail with a single solution. Our results showed that in 8 out of 1,010 tests, none of the four injected faults yielded a differential trail with a single solution. While this issue could be resolved in a real attack scenario by injecting additional faults, we excluded these 8 cases because the primary objective of this experiment is to statistically analyze the reduction of key candidates specifically when a deterministic trail is successfully formed. Consequently, we performed our statistical analysis on the 1,002 valid test cases where a unique solution was identified.

\subsection{Experimental Results}
\label{app:Experimental Results}

For the 1,002 valid experiments, we measured the average number of remaining key candidates at each recovery step. The results, expressed in the form of $2^k$, are as follows:

\begin{itemize}
    \item \textbf{Step 1 (Nibble Level):} This step involves guessing the last round key at the nibble level, with filtering performed at the first-to-last round. The experimental results indicated that the average number of remaining candidates per nibble was approximately $2^{2.18}$.
    \item \textbf{Step 2 (Group Level):} This step involves guessing the last round key at the group level, with filtering performed at the second-to-last round. Here, the average number of remaining candidates per group was approximately $2^{4.19}$.
    \item \textbf{Step 3 (Half Level):} This step involves guessing the last round key at the half level, with filtering performed at the third-to-last round. At this stage, the key candidate space reduced drastically, leaving an average of only $2^{0.42}$ candidates per half. This implies that in most cases, the key is uniquely determined or very few candidates remain.
    \item \textbf{Step 4 and Beyond (State Level):} This step involves guessing the last round key at the state level, with filtering performed at the fourth-to-last round and preceding rounds. The results showed that at Step 4 (Round 4), the average number of candidates converged to $2^{0.00}=1$. This demonstrates that the unique master key was successfully recovered in all 1,002 test cases upon reaching this stage.
\end{itemize}

This experimental data aligns closely with our theoretical calculations, demonstrating that our proposed attack framework is highly effective for key recovery in practice.

\end{document}